# Graphene on a hydrophobic substrate: Doping reduction and hysteresis suppression under ambient conditions.

*Myrsini Lafkioti, Benjamin Krauss, Timm Lohmann, Ute Zschieschang, Hagen Klauk, Klaus v. Klitzing and Jurgen H. Smet[*]*

Max Planck Institute for Solid State Research, Heisenbergstr.1, 70569 Stuttgart, Germany.

AUTHOR EMAIL ADDRESS J. Smet@fkf.mpg.de

TITLE RUNNING HEAD. Doping and hysteresis suppression in graphene

ABSTRACT **The intrinsic doping level of graphene prepared by mechanical exfoliation and standard lithography procedures on thermally oxidized silicon varies significantly and seems to depend strongly on processing details and the substrate morphology. Moreover, transport properties of such graphene devices suffer from hysteretic behavior under ambient conditions. The hysteresis presumably originates from dipolar adsorbates on the substrate or graphene surface. Here, we demonstrate that it is possible to reliably obtain low intrinsic doping levels and to strongly suppress hysteretic behavior even in ambient air by depositing graphene on top of a thin, hydrophobic self assembled layer of hexamethyldisilazane (HMDS). The HMDS serves as a reproducible template that prevents the adsorption of dipolar substances. It may also screen the influence of substrate deficiencies.**

KEYWORDS: Graphene, transport properties, hydrophobic, self-assembled layer, hexamethyldisilazane, hysteresis, doping.

Graphene has attracted considerable interest in recent years in view of the uncommon linear dispersion[1,2,3] for charge carriers and many desirable transport properties for graphene based electronics[4,5]. Among these are exceptional carrier density tunability including a reversal of the charge carrier polarity, high current densities[6], as well as equal or comparable mobilities for electrons and holes[7,8,9,10,11,12,13]. Graphene field effect transistors made by mechanically exfoliating graphene from graphite onto a thermally oxidized silicon substrate exhibit the highest quality up to this date among all explored approaches in which graphene is supported by a substrate[14,15,16,17,18,19,20,21]. Unfortunately, the characteristics of such field effect devices may vary widely. In particular the intrinsic doping level of as prepared devices as well as the mobility exhibit a large variance. Field effect characteristics also suffer from hysteretic behaviour, when measured under ambient conditions, as well as asymmetries between electron and hole transport[22,23,24,25,26,27]. Even though extensive systematic studies are still lacking and are also difficult to carry out, evidence accrues that morphology and deficiencies of the substrate, contamination during processing[28] as well as adsorbed molecules from ambient air[29] play a crucial role for these imperfections and the poor reproducibility of graphene devices. For instance very high mobilities were obtained in suspended graphene samples after current self-annealing[30,31], which was attributed to the absence of substrate effects and the successful removal of contaminations caused by the preparation procedures by the annealing process. Here we explore whether it is possible to also obtain reproducible characteristics for graphene supported by a substrate.

To identify a suitable approach, it is instrumental to summarize key experimental observations and theoretical considerations related to the intrinsic doping and hysteresis in graphene. The substrate surface and molecules adsorbed at this surface likely play a crucial role as they may impose their morphology on the deposited graphene[28,32]. The substrate surface quality itself depends on the morphology and the deficiencies of the $SiO_2$ top layer as well as on its chemical cleanliness. Various adsorbates can attach themselves to $SiO_2$. Hydroxyl groups (-OH) for instance couple to the dangling bonds of the Si on the surface and build a layer of silanol (SiOH) groups[33,34]. This silanol layer is very hydrophilic. Dipolar molecules can easily attach to the SiOH and contribute to the charge transfer, which results in doping of the graphene flake[35,36,37]. Most frequently p-doping is observed, which is believed to originate from adsorbed water molecules, possibly in combination with interactions between these molecules and the substrate[36,37].

The asymmetry in the conductivity and the hysteresis in the field effect may also originate from adsorbates[22,23,38], but both are still not fully understood. For example in the case of water, the most abundant dipolar adsorbate under ambient conditions, the doping and hysteresis mechanism are still controversially debated. Wehling et al. argued that only highly ordered $H_2O$ clusters[36] are able to act as dopants or doping from $H_2O$ molecules has to be mediated by defects in the $SiO_2$ substrate[37]. Such $H_2O$ molecules connect to the silanol groups on the surface. Lee et al.[39] concluded that the silanol groups themselves cause the hysteresis effect and adsorbates may just amplify it under ambient conditions. Leenaerts et al.[35] introduced the orientation of the water molecules as an important parameter controlling the doping effect of water. This work was based on DFT-calculations and did not require the presence of the substrate as a clustering template. The hysteresis in the field effect was also studied on carbon nanotubes[40,41,42,43]. Kim et al. for instance asserted that expanded clusters of water, which couple to the silanol groups of the substrate, surround the nanotube and cause the hysteretic behaviour. Mcgill et al.[42] have shown a reduction of hysteresis on SWNTs on a hydrophobic layer of octadecyltrichlorosilan (OTS).

According to our experience, the intrinsic doping level drops and hysteresis is suppressed or vanishes when placing the graphene under vacuum and pumping for an extended time. Heating the sample in vacuum to above 140 °C is very beneficial, but even without heating the hysteresis and doping level are reduced. This suggests that loosely bound species are the main culprits for hysteresis[22]. Strongly bound silanol groups or charge traps in the oxide would be stable even at elevated temperatures. An important observation is that most samples return to their initial state in terms of doping and hysteresis (within a tolerance of a few percent only) after a short time (<1 min) when exposing the flake back to air. This reversibility suggests that doping adsorbates preferentially attach to specific locations determined by the substrate specifics. A similar argument was invoked previously in ref. 22,44, and 45. Based on this information, we conclude that chemical hydrophobization of the substrate to remove and prevent the formation of silanol groups and thus the coupling of adsorbates, should provide a good solution to the venture of obtaining reproducible characteristics such as low intrinsic doping and weak hysteresis for graphene supported by a substrate. Here we show that a thin, hydrophobic self-assembled organic layer on top of the $SiO_2$ fulfils these requirements.

The substrates, which consist of an n+-Si wafer with a 300 nm thick thermal oxide, were prepared by the following procedure: The $SiO_2$ layer was cleaned in N-methyl-pyrrolidone, acetone and 2-propanol at 55°C. Subsequently, the substrate was treated in an $O_2$-plasma to remove organic residues. To hydrophobize the $SiO_2$ surface, the substrate was left in a HMDS[46] (hexamethyldisilazane/acetone) solution for 15-20h. The HMDS molecules (fig. 1b) form an ordered self assembled layer on the substrate. The long duration of exposure to HMDS was found to be crucial. Graphene is then deposited on the HMDS layer by micromechanical cleavage from HOPG[7] and identified by means of optical microscopy and Raman spectroscopy. To obtain a well defined geometry out of the graphene flakes, a Hall bar shape was patterned by using a PMMA mask defined by electron-beam lithography and an $O_2$ plasma etch. In a second e-beam lithography step, contacts were written and fabricated by evaporation of 3 nm Cr and 30 nm Au (fig. 1a and c). To characterize the sample doping, we studied the field effect at room temperature without and with annealing of the samples at ~140 °C for a time period of 1-2 hours. In addition quantum Hall effect (QHE) measurements were carried out at 1.6 K to assess the transport quality. For the sake of comparison, reference samples were prepared in the same fashion except that no HMDS layer was deposited prior to graphene exfoliation.

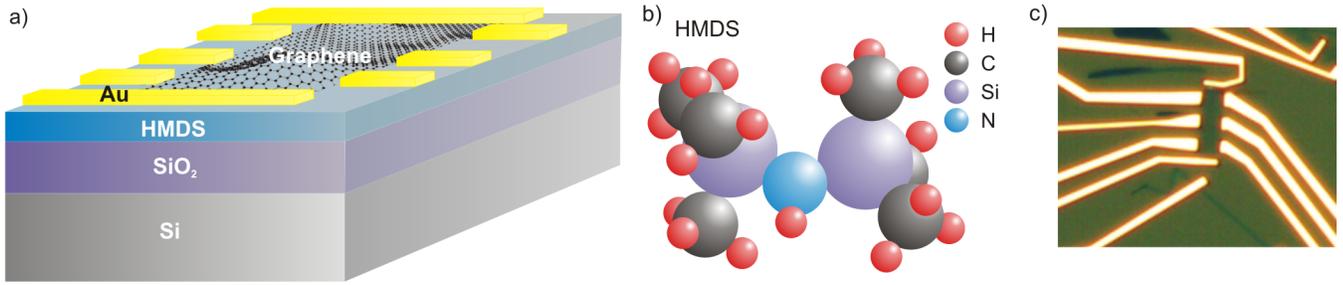

Figure 1: a) Schematic of the contacted graphene sample on top of an HMDS self-assembled layer. b) 3D-structure of the HMDS molecule. c) Optical image of the contacted graphene flake.

In Fig. 2a the mobility extracted from the field effect data and the intrinsic doping level are plotted for the reference graphene flakes prepared on bare $SiO_2$ with and without pumping and heat treatment (red circles and discs). The graphene samples deposited on a bare $SiO_2$ substrate exhibited charge neutrality at back gate voltages between +50V and +60V, which corresponds in our geometry ($0.7 \cdot 10^{11} cm^{-2}/V$) to a high p-doping level between 3.5 and $4.2 \cdot 10^{12} cm^{-2}$. Previous experience has shown however that this voltage varies strongly from flake to flake and seems to depend on processing details. The field effect curves for up and down sweeps of the back-gate voltage are depicted in Fig. 2b (solid and dotted red line respectively). A strong hysteresis is observed. It is attributed to dipolar adsorbates[22,43], the configuration of which changes upon sweeping. The different configurations produce an electric field that influences the charge carrier density in the sample. Under ambient conditions, the most probable candidate is water from the atmosphere. It has been demonstrated previously that vacuum annealing of the graphene samples (150°C, 1h) can help to remove adsorbates such as $H_2O$, $NO_x$, $CO_2$ and reduce both hysteresis[22] and the intrinsic doping level[29]. However, when exposing the sample back to air, approximately the same doping level is recovered and the hysteresis returns. This memory effect has been observed in several samples during our work and indicates that adsorbates responsible for doping return to the same amount on the coupling sites of the graphene flake. Moser et al.[44] have argued in a similar fashion as described in the introduction. Presumably defects such as edges, wrinkles, etc. serve as fixed docking sites on the graphene flake, which are not healed by heat treatment in vacuum.

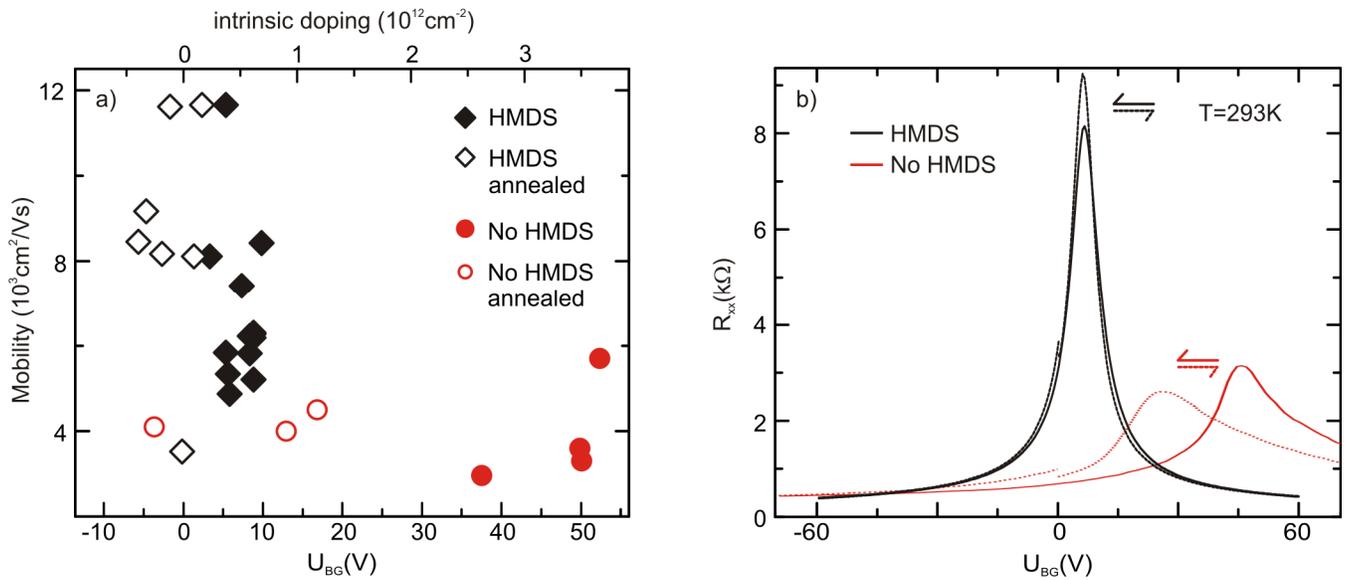

Figure 2: a) Mobility versus charge neutrality point of graphene deposited on bare $SiO_2$ (red circles and discs) and on HMDS (black full and empty diamonds), without (filled signs) and with (empty signs) annealing in vacuum (p~$10^{-6}$ mbar) at T = 140 °C for 1h. The mobility was determined at n = $1.25 \cdot 10^{12}$ $cm^{-2}$. The charge neutrality point for not annealed samples on bare $SiO_2$ is determined by the mean value of the charge neutrality points of both sweep directions as the exact doping cannot be measured due to the hysteresis. b) Field effect measurement at T = 293 K for graphene on HMDS (black curve) and for graphene on bare $SiO_2$ (red curve).

Fig. 2a also contains data points from a total of 13 graphene flakes without annealing and from 7 samples after annealing, all deposited on top of an HMDS self-assembled layer (black diamonds). Charge neutrality was reproducibly obtained at low back-gate voltages (<10 V) even without annealing. Field effect curves recorded during up and down sweeps of the back-gate voltage of a graphene flake on an HMDS treated substrate are plotted in Fig. 2b) (dotted and solid black lines respectively). Hysteresis has vanished nearly entirely, even under ambient conditions! Although significant scatter in the mobility remains, these HMDS treated samples on average exhibited higher charge carrier mobility. For the processing procedures described above and the HOPG starting material employed here, samples on bare $SiO_2$ typically had a mobility of ~4000 $cm^2$/Vs. Graphene prepared on HMDS-treated $SiO_2$ showed varying mobilities but values up to ~12.000$cm^2$/Vs were reached. Magnetotransport data recorded on a graphene sample deposited on top of a hydrophobic HMDS layer are plotted in Fig. 3.

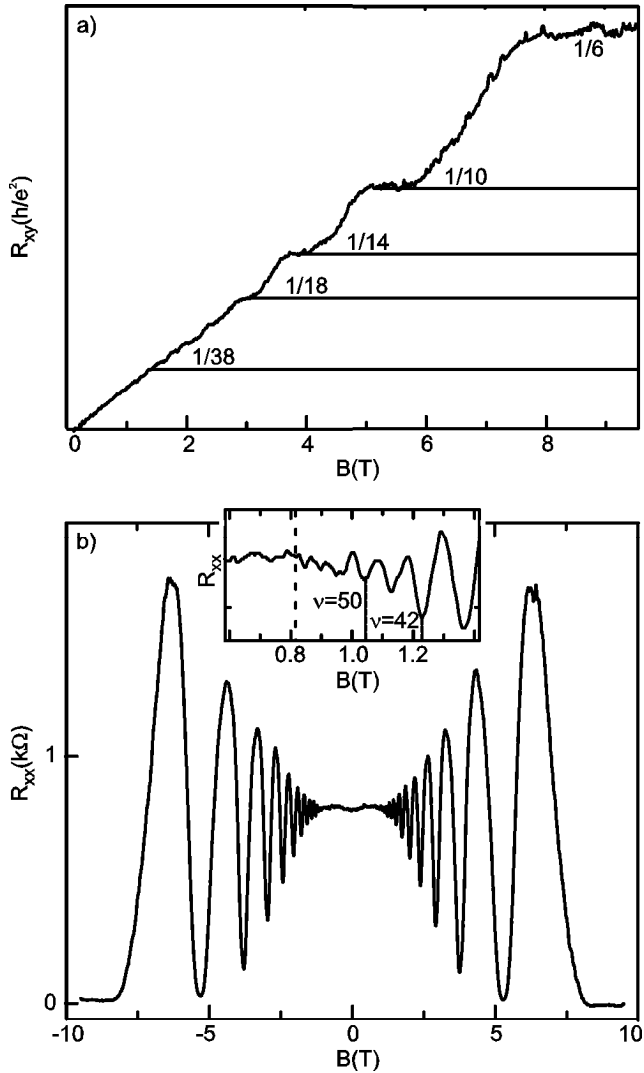

Figure 3: Hall effect measurement at T=1.6K on a graphene flake on HMDS (black curves) and on $SiO_2$ (red curves) (n = $1.25 \cdot 10^{12}$ $cm^{-2}$). a) $R_{xy}$ versus magnetic field, b) $R_{xx}$ versus magnetic field. Inset: magnification of the low field part of the positive field $R_{xx}$ measurement.

The Shubnikov de Haas (SdH) oscillations[8,9,12,13] exhibit good quality. They start at approximately 820 mT (inset to fig.3b). The longitudinal resistance is symmetric for both field directions and oscillation minima were observed up to a filling factor of 50. The extracted scattering time from the onset of the SdH oscillations was approximately 6.1 ps[30]. For a sample on untreated $SiO_2$ the scattering time was shorter by a factor of 2 or more. Despite this apparent improvement of the transport properties, the mobility values achieved are not able to compete with the mobilities reported on current annealed freestanding flakes[30,31]. This could be due to remaining effects of the template, resulting e.g. from defects in the deposited HMDS layer. At present insufficient statistics is available to conclude whether quantum Hall data is generally of higher quality in HMDS treated samples. We were however able to unequivocally establish that the main advantages of preparing graphene on HMDS are the reproducibly low intrinsic doping and the absence of hysteresis even under ambient conditions.

The drastic drop in the intrinsic doping level for graphene deposited on the HMDS self-assembled layer is attributed to its hydrophobic nature. The observed contact angle of water on the wafer serves as a measure of the hydrophobicity. For the HMDS layer we measured a contact angle of ~94°. A test measurement on a flake deposited on a self-assembled monolayer (SAM) of OTS with a smaller contact angle of ~73° resulted in comparable intrinsic doping (~0.5· $10^{12}$ cm$^{-2}$) compared to HMDS priming and reduced hysteresis[47]. Bare $SiO_2$ on the other hand exhibits a very small, with our setup not measurable, contact angle. It is hydrophilic since, without treatment, it is OH-terminated. Water molecules attach easily to the hydrogen of these silanol groups on the $SiO_2$ to form a thin water film. Molecules may also coalesce into clusters. The polar nature of water dopes the graphene layer and the arrangement into clusters may cause strain upon graphene deposition. HMDS apparently screens the flake from such influences. It likely displaces water molecules and clusters during its deposition as it can replace the OH groups on the substrate. Water molecules cannot attach or reorganize on the HMDS layer. The deposited graphene flake lies on a Si-C-H carpet, which forms a chemically well defined substrate with methyl groups that appear inert for the graphene flake. Loosely speaking, the HMDS layer may act as a kind of liquid surface on which the flake is floating. It shields the flake from defects of the substrate and reduces the adhesion to the surface.

In summary, a $Si/SiO_2$ substrate modified with a thin, hydrophobic organic template forms an excellent surface for the deposition of graphene. It inhibits polar adsorbates providing a chemically well defined and hydrophobic surface. The main merits are a reproducibly low intrinsic doping level largely independent from ambient conditions and processing details and a suppression of hysteretic behaviour in the field effect even under ambient conditions.

The authors thank K. Amsharov for fruitful discussions.

* J.Smet@fkf.mpg.de

[47] A SAM of pentadecylfluoro-octadecylphosphonic acid (CF3-$(CF_2)_6$-$(CH_2)_{11}$-$PO(OH)_2$) exhibited a contact angle of ~113°. However, we were not successful in fabricating graphene devices on this extremely hydrophobic surface. The SAM inhibits wetting of the surface with electron-beam lithography resist. Hence processing of contacts was not possible.